\documentclass[useAMS,usenatbib]{mnras}
\usepackage{graphicx}
\usepackage{url}
\usepackage{verbatim}
\usepackage{amsmath}
\usepackage{amssymb}

\newcommand{\mn}{\mnras}

\title[WD Pollution]{Polluting White Dwarfs with Perturbed Exo-Comets}
\author[Caiazzo \& Heyl]{Ilaria Caiazzo\thanks{E-mail:  ilariacaiazzo@phas.ubc.ca} and Jeremy S. Heyl \\
  Department of Physics and Astronomy, University of British
  Columbia, 6224 Agricultural Road, Vancouver, BC V6T 1Z1, Canada \\
}
\begin{document}
\date{Accepted 2017 April 27. Received 2017 April 26; in original form 2016 September 24}

\pagerange{\pageref{firstpage}--\pageref{lastpage}} \pubyear{2016}

\maketitle

\label{firstpage}

\begin{abstract}
  We present a model to account for the observed debris disks around young white dwarfs and the presence of metal-lines in their spectra.  Stellar evolution models predict that the mass-loss on the AGB will be pulsed; furthermore, observations indicate that the bulk of the mass-loss occurs on the AGB. In this case, if the progenitors of the white dwarfs had remnants of planetary formation like the Sun's Oort cloud or the Kuiper Belt and a planet lying within that cloud or nearby, we find that up to 2\% of the planetesimals will fall either into planet-crossing orbits or into chaotic regions after the mass-loss, depending on the location and mass of the planet (from Mars to Neptune).  This yields a sufficient mass of comets that can be scattered toward the star, form a debris disk and pollute the atmosphere.
\end{abstract}

\section{Introduction}
\label{sec:introduction}
The observation of metal absorption lines in the spectra of many white dwarfs in the last few decades has presented a theoretical challenge \citep{1980ApJ...242..195C,1982A&A...113L..13K,1983ApJ...272..660L,1986A&A...155..356Z}. The atmospheres of white dwarfs, whether they are hydrogen- or helium-dominated, possess short sedimentation timescales for heavy elements \citep{1992ApJS...82..505D}. For this reason, it is hard to explain the observed metal excess by the means of the stars' internal evolution only;  metals have to be continuously accreted onto the surface of the white dwarf from external objects. \citet{2010MNRAS.404.2123F} argue that the population of DZ white dwarfs (those with metal lines) is similar to DB white dwarfs (those that lack hydrogen); therefore, it is difficult to account for the lines by invoking interstellar accretion, which would bring hydrogen as well.  They suggest that the DZ white dwarfs accrete planetesimals. The lack of hydrogen in polluted white dwarf spectra \citep{2006ApJ...653..613J,2012MNRAS.424..333G}, even if with exceptions \citep{2017ApJ...836L...7X}, appears to indicate that water-rich planetesimals like comets are unlikely to be the pollutants. However, young white dwarfs are very luminous, so if comets were to be processed in a debris disk, they could lose most of the volatiles before accreting onto the white dwarf's atmosphere. In support of this picture, since 2007, many white dwarfs have been observed to possess an IR excess in their spectra~\citep{2007ApJ...657L..41S} that can be explained by the presence of a debris disk. In the case of the central star in the Helix Nebula, the mass of this disk is estimated to be $0.13 \mathrm{M}_\oplus$ \citep{2007ApJ...657L..41S}.  It has been estimated that about 15\% of young white dwarfs possesses a debris disk~\citep{2011AJ....142...75C}.

The formation of the debris disk represents another puzzling problem. In the commonly accepted scenario, small rocky bodies, like asteroids, are accreted onto the white dwarf thanks to their interaction with planets that have survived the post-main-sequence evolution of the star. This scenario, however, is hard to implement in reality because, during the last stages of stellar evolution, the inner solar system will evolve adiabatically \citep{2013MNRAS.435.2416V,1998ApJ...501..357P,2010MNRAS.404.2123F}, which means that the orbits of all the planets and asteroids will only scale up and this makes it difficult for them to be tidally disrupted. Models have also been developed in which the metals are accreted from bodies coming from beyond the ice line~\citep{1986ApJ...302..462A,2015MNRAS.448..188S}. In these models, extremely strong stellar winds or large white dwarf kicks have been proposed to explain how the orbits get disrupted. A review of the models proposed can be found in \citet{2016RSOS....350571V}.

\citet{2011MNRAS.414..930B} hypothesize the presence of a planet close in semi-major axis to a Kuiper belt-like cometary population to explain how comets' orbits can be perturbed at the end of a star's post-main sequence. We develop this idea further, showing how the number of comets entering the chaotic regions close to the planet is enhanced when one considers realistic mass-loss timescales on the asymptotic giant branch. Furthermore, we show that if the planet's orbit lays further out in semi-major axis, closer to the Oort cloud, the non-adiabatic evolution of objects in this region increases the number of comets that end up on unstable orbits by up to almost two orders of magnitude (0.07\%-1.85\%). 

During the post-main sequence stages of a star's evolution, as the star loses a large fraction of its mass, the semi-major axis of the objects in the inner star system expand adiabatically, because the mass-loss timescale is large compared to their orbital periods. During the adiabatic expansion, the product $Ma$ between the star's mass and the semi-major axis of the orbiting object remains constant because it is an adiabatic invariant. This means that all the orbits only scale up during the mass-loss phase; however, chaotic regions around planets can grow as the relative influence of the planets compared to the star increases, leading a small number of planetesimals to end up on unstable orbits.  For objects in the outer stellar system, this approximation does not hold because their orbital frequencies and the star's relative mass-loss rate can be comparable. In this region, orbits don't scale up, but they evolve impulsively and their evolution depends on the true anomaly of the objects at the moment of the mass-loss (see \S~\ref{sec:evol-comet-popul}). If a planet is close enough, comets can still fall into its chaotic regions because the regions themselves grow, but many more will end up on planet-crossing orbits owing to non-adiabatic evolution.

The position of the adiabatic-to-impulsive transition region depends on the star's mass-loss rate. In \S~\ref{sec:mass-loss-models}, we employ MESA stellar evolution code~\citep{2011ApJS..192....3P} to obtain models for stellar evolution with different mass-loss rates on the red giant branch (RGB) and on the asymptotic giant branch (AGB) to see how the different evolution paths affect the expansion of orbiting objects' semi-major axis. We show that, for stellar evolution models in which the bulk of the mass-loss happens on the RGB, the orbital evolution is impulsive only for objects with semi-major axis $a \gtrsim 10^5$ AU.  On the other hand, when the star loses its mass mainly on the AGB, the mass-loss is rapid enough that objects in the same position as our Oort cloud (between $10^3$ and $10^5$ AU) evolve impulsively. In order to simulate the evolution of a large number of comets with different mass-loss models, we use an interpolation formula (\S~\ref{sec:evol-comet-popul}) to describe the adiabatic-to-impulsive transition region. Investigation the full two-body point-mass problem is possible, even if more demanding computationally \citep{2011MNRAS.417.2104V}.  In \S~\ref{sec:role-planet} we argue that, if there is a planet in the same region as the Oort cloud (or the Kuiper belt), it can scatter comets to eccentricities high enough that they can reach close to the white dwarf and ultimately accrete onto it. In \S~\ref{sec:unstable-planetesimals} we describe different possible scenarios in which cometary reservoirs like our Kuiper belt and our Oort cloud can provide the amount of planetesimals needed to account for white dwarfs' metal lines and debris disks. 

\section{Mass-Loss Models}
\label{sec:mass-loss-models}

We use a suite of mass-loss models calculated with MESA
\citep{2011ApJS..192....3P} and calibrated for the mass-loss history of the stars in 47~Tucanae for which we have direct evidence of the epoch and extent of mass-loss \citep{Heyl14massloss,2016arXiv160505740P}. These recently evolved stars had initial masses of 0.9 solar masses and resulted in 0.53 solar mass white dwarfs \citep{2009ApJ...705..408K}, so the total fractional mass-loss is about forty percent.  More massive main sequence stars will lose a larger fraction of their mass, so our model provides a lower limit of the effects of mass-loss on the orbital distribution of the cometary bodies.

For the MESA models, we used SVN revision 5456 and started with the model {\tt 1M\_pre\_ms\_to\_wd} in the test suite.  We changed the parameters {\tt initial\_mass} and {\tt initial\_z} of the star (respectively, the star's mass and metal content at the beginning of the simulation) and adjusted the parameter {\tt log\_L\_lower\_limit} to $-6$, which means that we set the stopping condition for the simulations when the star's luminosity is less than $10^{-6} L_\odot$, so that the simulations run well into the white dwarf cooling regime. For the mass-loss on the red-giant branch we used the \citet{1975MSRSL...8..369R} value,
\begin{equation}
{\dot M}_R =   4 \times 10^{-13} \eta_\mathrm{RGB}  \frac{L}{L_\odot} \frac{R}{R_\odot}
\frac{M_\odot}{M} \,\,\,\,\,\,\,\,\,\,\,\,\,[{M_\odot} \,{\mathrm{ yr}}^{-1}]
\label{eq:1}
\end{equation}
where $M$, $R$ and $L$ are the star's mass, radius and luminosity respectively. $\eta_\mathrm{RGB}$ is a constant that determines the efficiency of the mass-loss on the RGB. On the asymptotic-giant branch we use the \citet{1995A&A...297..727B} formula
{\small
\begin{equation}
{\dot M}_B = 1.93 \times 10^{-21} \eta_\mathrm{AGB} \left (\frac{M}{M_\odot} \right)^{-3.1} \left (\frac{L}{L_\odot} \right)^{3.7} \frac{R}{R_\odot} \,\,\,\,[{M_\odot} \,{\mathrm{ yr}}^{-1}].
\label{eq:2}
\end{equation}
}
where $\eta_\mathrm{AGB}$ is a constant that determines the efficiency of the mass-loss on the AGB.

We reduced the value of $\eta_\mathrm{RGB}$ to 0.1, 0.3 and 0.6 (from the default of 0.7) in the range of \citet{2012MNRAS.419.2077M} and we used higher values of $\eta_\mathrm{AGB}$ as outlined in Table~\ref{tab:windparam}. These parameters yield a white dwarf whose mass is about 0.53 solar masses \citep{1988ARA&A..26..199R,2004A&A...420..515M,2009ApJ...705..408K} from the 0.9 solar mass progenitor.  The MESA models are consistent with the observed $M_V$ of the tip of the TP-AGB \citep{2005A&A...441.1117L} which is sensitive to the mass of the resulting WD and with the observed cooling curve of the white dwarfs \citep{Gold15wd}.

Table~\ref{tab:windparam} summarizes the results of the various wind
models using MESA.  Essentially, the first wind parameter,
$\eta_\mathrm{RGB}$, can be tuned to change the amount of mass-loss on
the red-giant branch and, therefore, the mass of the horizontal branch (HB)
stars.  The value of the second parameter, $\eta_\mathrm{AGB}$, does not
affect the final mass of the white dwarf but rather determines whether
the mass-loss on the AGB will be more or less pulsed and the thickness
of the residual hydrogen layer on the white dwarf \citep{Gold15wd}. Since the final mass is determined by the evolution on the RGB, the amount of mass that is lost on the AGB is fixed. If the efficiency of the mass-loss on the AGB is poor ($\eta_\mathrm{AGB}$ is small) then little mass is lost over time, until the system becomes unstable and the bulk of the mass is lost in outbursts. Thus, a low value of $\eta_\mathrm{AGB}$ corresponds to a pulsed mass-loss.
\begin{table}
  \caption{Wind Parameters for the Various MESA Models.}
  \centering
\label{tab:windparam}
\begin{tabular}{ccccc}
\hline
$\eta_\mathrm{RGB}$ & $\eta_\mathrm{AGB}$ & $M_\mathrm{HB}$ & $M_\mathrm{WD}$ & Color in Fig.~\ref{fig:mass-mdot}   \\
                 &                    &  [$M_\odot$] & [$M_\odot$] & \\
\hline
0.1  & 0.7 & 0.86 & 0.54 & blue  \\
0.1  & 0.52 & 0.86 & 0.54 & cyan \\ 
0.1  & 0.21 & 0.86 & 0.55 & magenta\\
0.3  & 0.5 & 0.78 & 0.53 & green \\
0.6  & 0.7 & 0.63 & 0.52 & red \\ 
0.6  & 0.1 & 0.63 & 0.53 & ochre
\end{tabular}
\end{table}

\begin{figure}
\includegraphics[width=\columnwidth]{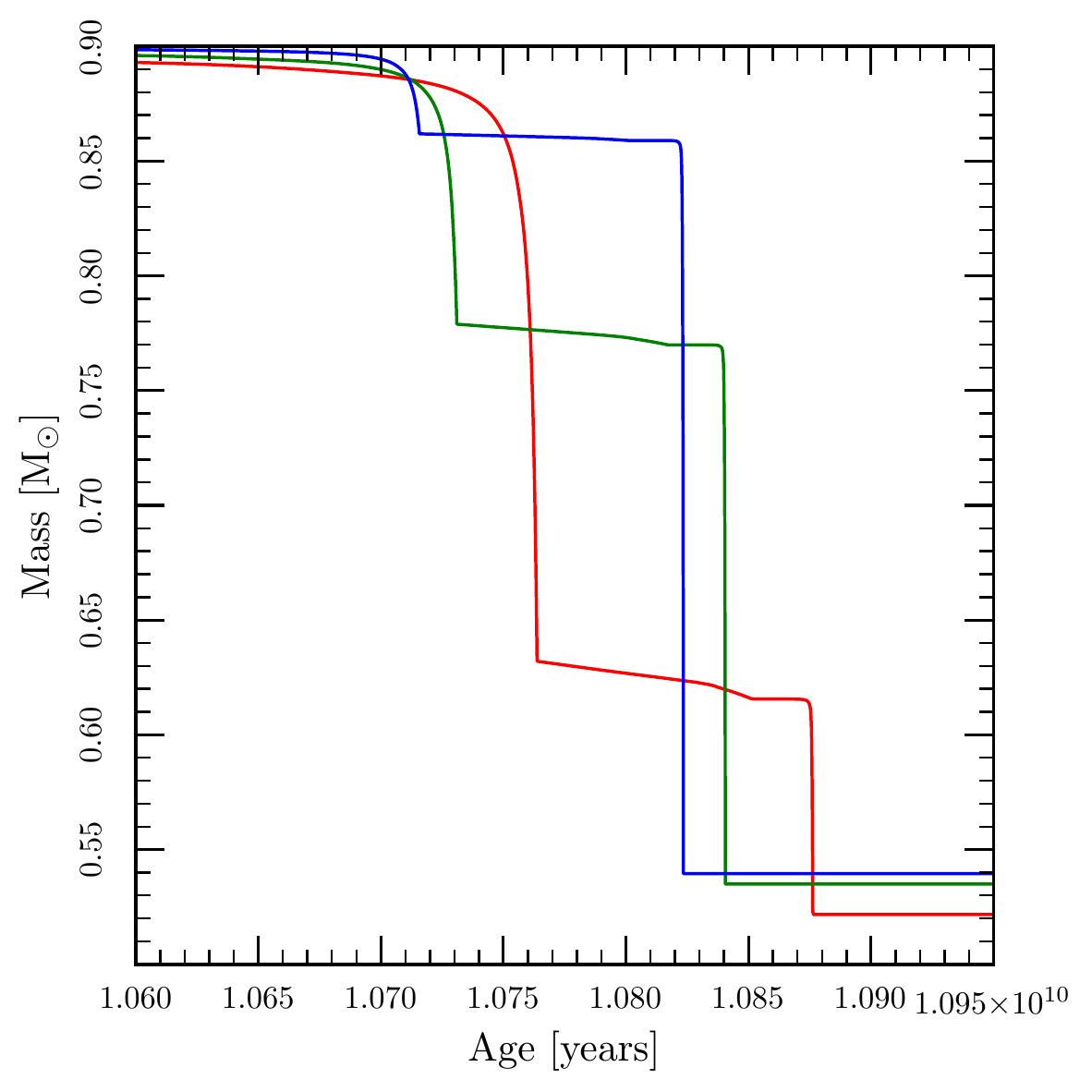}
\caption{Mass of a star that will become a modern-day white dwarf
  as a function of its age for the MESA models.  The first epoch of mass-loss
  in the model is as a red giant and the second is as an asymptotic giant star.
  The curves trace the mass-loss for $\eta_\mathrm{RGB}=0.1$ (blue), $0.3$ (green) and
  $0.6$ (red) from top to bottom with the other quantities given in
  Table~\ref{tab:windparam}.}
\label{fig:time-mass}
\end{figure}

\begin{figure}
\includegraphics[width=\columnwidth]{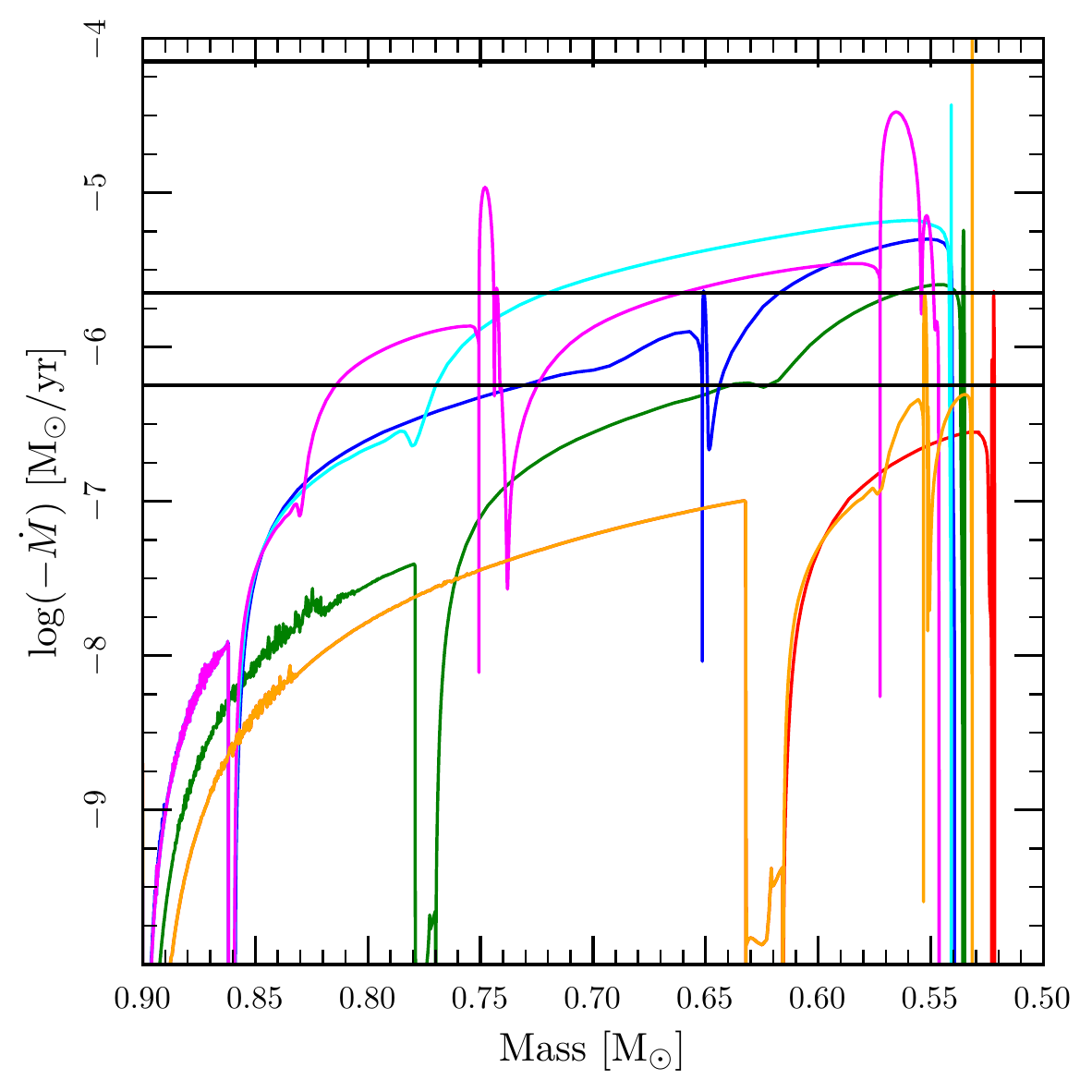}
\caption{Mass-Loss rate of a star that will become a modern-day white
  dwarf as a function of its mass for the MESA models.  The first
  epoch of mass-loss in the model is as a red giant and the second is
  as an asymptotic giant star.  The values of $\eta$ are the same as
  in Fig.~\ref{fig:time-mass} (for red, green and blue).  Two additional
  models are depicted
  with $\eta_\mathrm{RGB}=0.1$ and $\eta_\mathrm{AGB}=0.21$ (magenta)
  and 0.516 (cyan) and one has $\eta_\mathrm{RGB}=0.6$ and
  $\eta_\mathrm{AGB}=0.1$ (ochre). Black solid lines, from top to bottom:
  orbital angular frequencies at the inner, inner to outer, outer edge
  of the Oort Cloud.}
\label{fig:mass-mdot}
\end{figure}

\citet{Heyl14massloss} found that the bulk of the mass-loss of the
stars that are now becoming white dwarfs in 47~Tucanae occurred on the
AGB, and \citet{2016arXiv160505740P} found that the mass of the
horizontal branch stars in the same cluster is greater than
0.81M$_\odot$. Our focus will be on the models with the largest HB mass
of 0.86M$_\odot$ and $\eta_\mathrm{RGB}=0.1$. To understand the role
of how the mass is lost on the AGB, we also consider models of
$\eta_\mathrm{AGB}=0.21$ and 0.516.  These models appear identical to
the blue model in Fig.~\ref{fig:time-mass}, but the lower value of
$\eta_\mathrm{AGB}$ results in a more pulsed mass-loss during the AGB, as
depicted in Fig.~\ref{fig:mass-mdot}.  We also examine two models with
the bulk of the mass-loss on the RGB resulting in a horizontal branch star of
0.63~solar masses, one with smooth mass-loss and one more pulsed.

Strong evidence for a pulsed mass-loss during the final stages of the AGB evolution is provided by observations of planetary nebulae. A recent analysis of proto-planetary nebulae and planetary nebulae observations from the \textit{Hubble Space Telescope} and the \textit{Spitzer Space Telescope}, found that about eight percent of these objects present rings and arcs around them \citep{2016MNRAS.462..610R}. These rings are commonly interpreted as quasi-periodic enhancements in the mass-loss rate and the time lapse between them varies between 100 to 2000 years.  \citet{2001AJ....121..354B} estimate for 
NGC 6543 (the Cat's Eye Nebula) that the mass of each ring is approximately $0.01~\mathrm{M}_\odot$ and that the star lost about a tenth of a solar mass over about 10,000 years.  These pulses are even more pronounced than the ones we obtain in our models. 

\section{Evolution of the Cometary Population}
\label{sec:evol-comet-popul}
In Fig.~\ref{fig:mass-mdot}, the mass-loss rate for each model is plotted as
a function of the instantaneous mass of the star. The increase in the
mass-loss rate on the leftmost side of each curve corresponds to the
RGB. As the stars reach the HB, the mass-loss rate decreases
dramatically; this phase is evident in Fig.~\ref{fig:mass-mdot} as a dip in the mass-loss.  When a star leaves the HB and rises up the AGB, it loses mass quickly; however, for smaller values of $\eta_\mathrm{AGB}$ and larger values of the mass of the HB star, the stars reach higher peak mass-loss rates. In order to understand the effect of the mass-loss on the evolution of objects orbiting
around the star, we plotted on the same figure (black solid
lines), superimposed from top to bottom, the orbital angular frequencies for objects at the inner edge of the inner Oort Cloud around the Sun (2,000~AU), at the transition between the inner and outer Oort Cloud (20,000~AU) and at
the outer edge of the outer Oort cloud \citep[50,000 AU;][]{2004ASPC..323..371D}. If the mass of the star changes little
over the course of an orbit, the orbits simply expand as the product $Ma$
of the mass and the semi-major axis is an adiabatic invariant.
Furthermore, the shape of the orbit remains constant in this process
\citep{Heyl07kickbin}.  On the other hand, if the mass changes
substantially over the orbital period, the mass-loss is impulsive
from the point of view of the comet, so the shape of its orbit will change.  This figure gives us a sense of the fate of the Oort cloud, and, if we consider
that the progenitors of the white dwarfs for which we see metal lines
had cometary clouds similar to that around the Sun, it provides a mechanism to
disturb the Oort cloud and possibly deliver cometary material inward.

First, we see that, for models where the horizontal branch star has a small value, such as 0.63M$_\odot$, the mass-loss rate on the AGB never exceeds the orbital frequency at the outer edge of the solar system, except for very short periods of time, so the entire cometary system will expand adiabatically during the RGB and AGB, yielding a slight increase in cometary infall due to passing stars as in our solar system.  On the other hand, for larger horizontal branch masses, we have a mixture of adiabatic and impulsive evolution within the cometary system.  In particular, as the value of $\eta_\mathrm{AGB}$ decreases and $M_\mathrm{HB}$ increases we have a large fraction of the cometary system evolving impulsively. 

To examine this further, we develop a simplified model to track the evolution of an initial distribution of cometary bodies that are evenly spaced in $\log a$ \citep{2004ASPC..323..371D}. Instead of following the evolution of each orbit, we use the fact that $Ma$ is an adiabatic invariant to track the evolution of comets with short period orbits, while we keep the true anomaly fixed for comets with long period orbits (outside of the adiabatic regime) as we change the star's mass (we call this regime impulsive).

To estimate the impulsive evolution we use the relation between the semi-major axis and the energy
\begin{equation}
  \frac{G M \mu}{2 a} = \frac{GM\mu}{r} - \frac{1}{2}\mu v^2.
  \label{eq:5}
\end{equation}
where $M$ is the mass of the star and $a$, $\mu$ and $v$ are the semi-major axis, the mass and the orbital velocity of the comet.
This relation holds true both before and after the mass-loss, but of course the value of the energy changes due to the mass-loss.  Solving for $a$ yields
\begin{equation}
  a = \left ( \frac{2}{r} - \frac{ v^2}{G M} \right )^{-1}
  = \frac{r}{2} \left ( 1 - \frac{r v^2}{2 G M} \right )^{-1}
  \label{eq:6}
\end{equation}
and we take everything fixed on the right-hand side except for the mass $M$ to get 
\begin{equation}
  \left . \frac{d \ln a}{d \ln M} \right |_\mathrm{impulsive}
  = -\frac{v^2 a}{G M} = 1 - \frac{2a}{r} =1 - 2 \frac{1+e\cos\theta_0}{1-e^2}
  \label{eq:7}
\end{equation}
where $\theta_0$ is the true anomaly along the orbit when the mass-loss
occurs and $e$ is the eccentricity.  As a check, we look at the average behaviour over an ensemble of
comets in a particular orbit. The time average of $\cos\theta$ is:
\begin{align}
  \left  \langle \cos\theta \right \rangle_\mathrm{adiabatic}  &=
  \frac{1}{P} \int_0^P \cos\theta dt \nonumber \\
  &= \frac{\left( 1-e^2 \right)^{3/2}}{2\pi} \int_0^\pi \frac{\cos\theta}{(1+e\cos\theta)^2} d\theta = -e
  \label{eq:8}
\end{align}
Where we used the method of residues. Substituting in Eq.~(\ref{eq:7}):
\begin{equation}
\left \langle   \frac{d \ln a}{d \ln M} \right \rangle_\mathrm{impulsive} = -1
\label{eq:9}
\end{equation}
We find that on average, the comets follow an adiabatic evolution, as 
expected.

In general, the change of $a$ in the impulsive limit depends on the eccentricity
and on the true anomaly at the moment of the impulsive mass-loss. We must then
find an expression for the impulsive evolution of the eccentricity as well.
Let us start with the value of the eccentricity
\begin{equation}
   1-e^2 = \frac{2 L^2}{G^2 M^2 \mu^3} \left ( \frac{G M \mu}{r} - \frac{1}{2} \mu v^2 \right )
\label{eq:10}
\end{equation}
where $L$ is the orbital angular momentum. Again we take everything fixed on the right-hand side except for the mass
$M$ to get
\begin{align}
\left . \frac{d \ln \left( 1-e^2\right )}{d\ln M} \right |_\mathrm{impulsive} &=
    -2 + \frac{2GM}{2GM - rv^2}  \nonumber \\ &= -2 + \frac{2 a}{r} = -2 + 2\frac{1+e\cos\theta_0}{1-e^2}
\label{eq:11}
\end{align}
Using the results from Eq.~(\ref{eq:8}), we obtain
the orbit-averaged value of
\begin{equation}
\left \langle \frac{d \ln \left( 1-e^2\right )}{d\ln M}  \right \rangle_\mathrm{impulsive} =
  - 2 + 2\frac{1-e^2}{1-e^2} = 0
\label{eq:12}
\end{equation}
Again, we find that the average behavior corresponds to an adiabatic
expansion.  Eq.~(\ref{eq:7}) and Eq.~(\ref{eq:11}) are equivalent to
those derived by \citet{2013MNRAS.435.2416V} (their equations 24 and
25), but in our calculations we keep the true anomaly fixed as we change the star's mass. We do not consider the effects of white dwarf
kicks, so we do not have to keep track of the orientation of the
orbit.  An important subtlety of Eq.~(\ref{eq:11}) is that the bulk of
the comets move to slightly smaller eccentricities and that this is
balanced by a fraction whose eccentricities increases by a larger
amount.

We treat the transition from the adiabatic to the impulsive regime using a formalism inspired by the Landau-Zener formula \citep{landau1932theorie,1932RSPSA.137..696Z,doi:10.1021/jp040627u}. We estimate the probability that the orbit fails to evolve adiabatically as the value of $M$ changes to be
\begin{equation}
  P = \exp \left [ -\frac{\pi}{2}  \left ( \frac{\Omega M}{\dot M} \right )^3  \right ]
  \label{eq:3}
\end{equation}
where $\Omega$ is the orbital angular frequency.
We take the evolution of the semi-major axis of an orbit to be given by the
following equation
\begin{equation}
  \frac{d\ln a}{d\ln M} = - ( 1 - P ) + P \left . \frac{d \ln a}{d \ln M} \right |_\mathrm{impulsive} 
  \label{eq:4}
\end{equation}
so when the fractional change in mass during an orbit is small, the semi-major axis increases adiabatically as $1/M$.  On the other hand, when the change is large, $a$ follows the impulsive evolution.

\begin{figure}
  \includegraphics[width=0.48\textwidth]{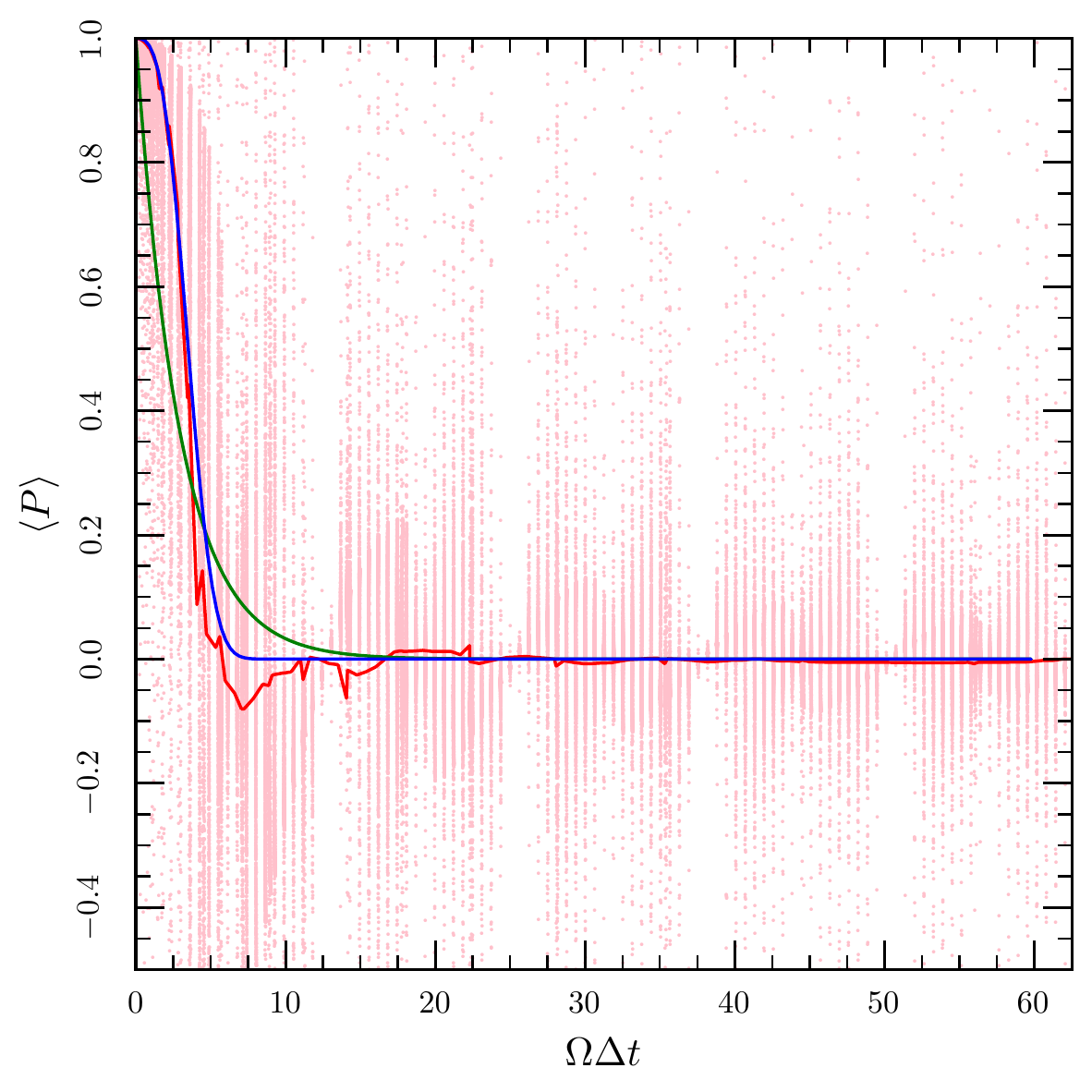}
  \caption{The pink dots show the mean value of $\cos\theta$ over an interval of time compared to the values in the impulsive limit $\cos\theta_0$ and the adiabatic limit $-e$.  The red curve is the median over a sample of the integrations.  The blue curve is the Landau-Zener interpolation that we use (Eq.~\ref{eq:3}), and the green curve represents the same function with an exponent of unity.}
  \label{fig:iwd2b}
\end{figure}

Of course, this is a drastic simplification of the dynamics, but it will give a sense of the resulting distribution, at least on average.  We motivate the validity of the Landau-Zener formula (Eq.~\ref{eq:3}) by calculating the mean of $\cos\theta$ over an interval of time $\Delta t$ and comparing it to the mean over many orbits, the adiabatic limit of $-e$, and to the impulsive value $\cos\theta_0$.  To start the calculation, we first solve for $P$ in Eq.~\ref{eq:4}
\begin{eqnarray}
P &=&  \frac{\left . \frac{d\ln a}{d\ln M} \right |_\mathrm{adiabatic} -  \frac{d\ln a}{d\ln M} }{
 \left . \frac{d\ln a}{d\ln M} \right |_\mathrm{adiabatic} - \left . \frac{d\ln a}{d\ln M} \right |_\mathrm{impulsive} } \\
 &=& \frac{\left \langle \cos\theta \right \rangle_\mathrm{adiabatic} -
 \cos\theta }{\left \langle \cos\theta \right \rangle_\mathrm{adiabatic} -
 \left .\cos\theta \right  | _\mathrm{impulsive}} \\
 &=& \frac{-e -  \cos\theta }{-e -
  \cos\theta_0 } = \frac{e +  \cos\theta }{e +
  \cos\theta_0 }.
\end{eqnarray}
The adiabatic value of $d\ln a/d\ln M$ simply corresponds to the average over many orbits with the parameters kept fixed, while the impulsive value is the instantaneous value when the mass changes.   We would like to obtain an estimate of the appropriate average value of $d\ln a/d\ln M$ to use when the mass-loss rate is intermediate between these two extremes.   To do this, we calculate the time average of $\cos\theta$ for an interval of time and find the appropriate average value of $P$, $\langle P \rangle$, to use in the orbital evolution.  What is important is the number of orbits over which the averaging is calculated or the total change in the mean anomaly $\Omega \Delta t$. For each of these averages we hold the mass of the star fixed, so the orbital parameters are fixed as well. When we integrate Eq.~\ref{eq:4}, over each step the mass changes by less than one part per thousand; therefore, this is a reasonable approximation. For each time average, we choose an initial mean anomaly randomly from a uniform distribution and an eccentricity from a distribution uniform in $e^2$.   

The pink dots in Fig.~\ref{fig:iwd2b} depict the results of the individual calculations, where we varied the initial values of the mean anomaly, the eccentricity and the duration of time average ($\Delta t$).  When the average is over an integral number of periods (that is, when $\Omega \Delta t \approx 2 \pi, 4\pi \ldots$), the average of the cosine yields the adiabatic value, so $\langle P \rangle$ vanishes.  This is apparent in the beat pattern in Fig.~\ref{fig:iwd2b}.  Furthermore, the average approaches the adiabatic value for large values of $\Omega \Delta t$ so the magnitude of $\langle P\rangle$ decreases.

In our simulations, as we evolve the orbital parameters for each object using Eq.~\ref{eq:4}, we do not wish to keep track of the phase of the orbits themselves, but we want to approximate the behavior using an appropriate mean value of $P$, so we calculate the median value of $P$ over the time averages for a particular duration $\Omega \Delta t$. The red curve depicts these median values. The blue and green curves show the Landau-Zener interpolation formula with an exponent of three and one respectively.  We find that the expression in Eq.~\ref{eq:3} (blue curve) yields the best interpolation for the median behavior. For this reason, we employ the Landau-Zener formula with the exponent of three in this paper's simulations. As a check, we also ran the same simulations for an exponent of one and we found a similar number of perturbed comets.

In the calculations shown in Fig.~\ref{fig:iwd2b}, we see the ambiguity in the boundary between the adiabatic limit and the impulsive limit.  We integrated over various numbers of orbits in time, {\em i.e.} $\Omega \Delta t$, to get a sense of the form of the best interpolation.  However, the relationship between $\Omega\Delta t$ and $\Omega M/{\dot M}$ is still arbitrary, as it reflects the arbitrariness of the $\pi/2$ prefactor. We find that the simulations are on average halfway between the two limits for $\Omega \Delta t \approx \pi$, which yields $\Omega \Delta t \approx 4.6 \, \Omega M/{\dot M}$ when we fit the behavior in Fig.~\ref{fig:iwd2b} with Eq.~\ref{eq:3}. Our stars experience a wide range of mass-loss rates, and we also simulate orbits with a wide range of $\Omega$, so we argue that our results are robust against changes in the prefactor.

In the next section, we show the results of simulations in which we employ this formalism to evolve the orbits of a sample of comets with an initially uniform distribution in $\log a$ \citep{2004ASPC..323..371D} and a distribution of eccentricities which is proportional to the eccentricity itself \citep{2015MNRAS.448..188S}.  For each comet, we pick a random number between 0 and $\pi$ for the mean anomaly $M$.  We use this same mean anomaly for all of the impulsive evolution.  For each step of the mass-loss we calculate the corresponding true anomaly using Kepler's equations:
\begin{align}
  &M = E - e\sin E \label{eq:anomaly1} \\
  &\tan\frac{\theta}{2} = \sqrt{\frac{1+e}{1-e}} \tan\frac{E}{2} \label{eq:anomaly2}
\end{align}
where $E$ is the eccentric anomaly and $\theta$ is the true anomaly. 

\section{The Role of the Planet}
\label{sec:role-planet}
\begin{figure*}
  \includegraphics[width=0.46\textwidth]{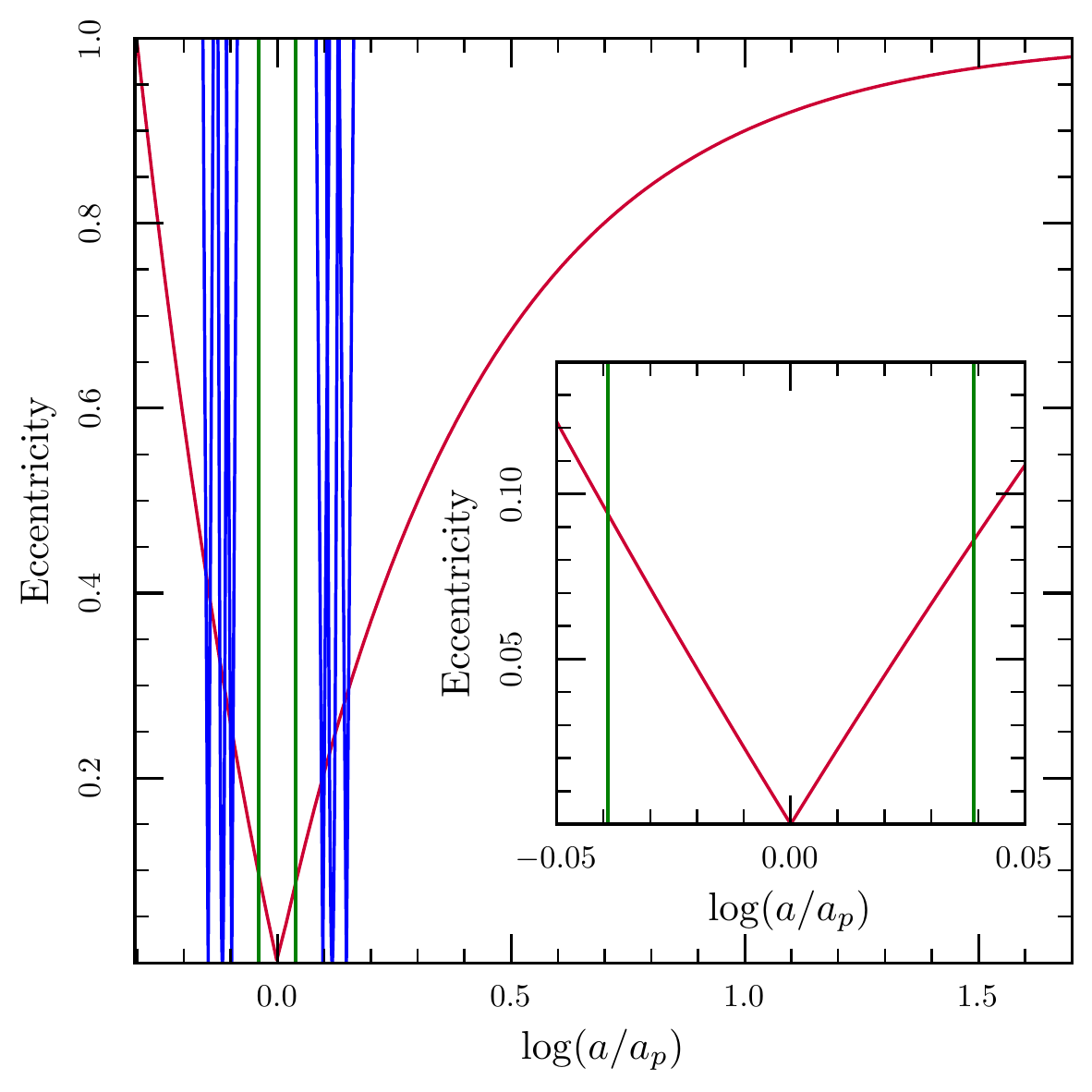}
  \includegraphics[width=0.46\textwidth]{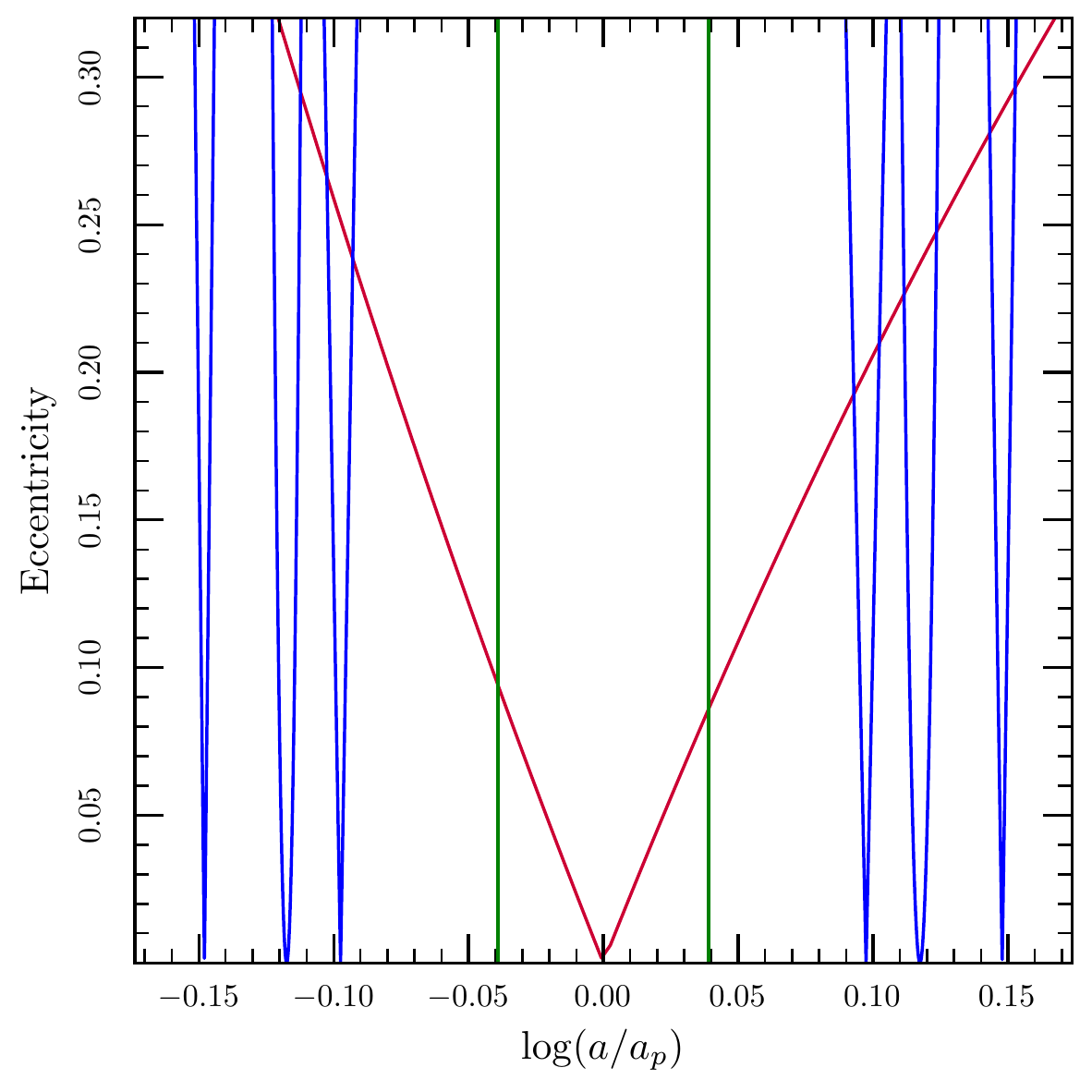}
  \caption{ The phase space of orbits of planetesimals near a
    Neptune-mass planet. The green lines delimit what we call the Wisdom-Duncan region in the text; orbits in this region are unstable.  Orbits above each of the blue
    curves lie within the 5:3, 3:2, 7:5 and 5:7, 2:3 and 3:5
    mean motion resonances from left to right where we used the formulae of \citet{solarsystemdynamics}
    to draw the boundaries.  The extension
    to large eccentricity is only illustrative as the expansion of the
    disturbing function is not accurate in this region. Orbits above the red line cross the orbit of the planet.  The orbits in the resonances (blue lines) are phase-protected from close encounters with the planet. The insert and the right panel show a zoom of the same plot.}
  \label{fig:res_diagram}
\end{figure*}

In our simulations, we hypothesize the presence of a planet, between Mars and Neptune in size, with a semi-major axis $a_p$ close to the Oort Cloud \citep[e.g. Planet 9,][]{2016AJ....151...22B,2016ApJ...824L..22M}. Because the mass of the planet is small (less than that of Saturn), we expect the overlap of the major mean-motion resonances (e.g. 3:2 with 5:3 and 7:5) to be small as well \citep{2016ApJ...824L..22M}.  The minimum mass of Mars is chosen so that the scattering rate is sufficiently large to drive planetesimals inward \citep{1995Icar..118..322M,2012A&A...548A.104B}. Before the onset of the mass-loss, we assume the star system to be relaxed. Thus, we
expect the comets to lie both within the low-order resonances depicted in Fig.~\ref{fig:res_diagram} and in the region where they do not cross the planet's orbit.
Additionally, we assume no comets in
the region with semi-major axes close to that of the planet
\citep{1980AJ.....85.1122W,1989Icar...82..402D}:
\begin{equation}
  \frac{|a-a_p|}{a_p} < 1.49 \left ( \frac{m_p}{m_p+M} \right )^{2/7}.
  \label{eq:wisdom-duncan}
\end{equation}
for convenience, we will call this chaotic region the Wisdom-Duncan region, as its expression was derived analytically by \citet{1980AJ.....85.1122W} and we chose the factor 1.49 from posterior numerical refinements, such the ones by \citet{1989Icar...82..402D}. This region plays a dominant role only for comets with low eccentricity.  We assume that the libration timescale \citep[a few Myr,][]{2016ApJ...824L..22M} is much longer than the mass-loss timescale so we ignore the direct effect of the planet on the orbits during the mass-loss itself.

Fig.~\ref{fig:res_diagram} depicts the width of the Wisdom-Duncan region near to the planetary orbit (green lines), the region in which the orbits cross the orbit of the planet (above the red lines) and the resonance regions (between blue lines).  We focus on the first-order 3:2 and 2:3 resonances and the nearby second-order 5:3, 7:5, 5:7 and 3:5. The extent of these resonances regions is given by \citep[see Eq. 8.58 of][]{solarsystemdynamics}
\begin{equation}
\frac{|a-\alpha a_p|}{a_p} < \left ( \frac{16}{3} \frac{m_p}{m_p+M} \alpha f_d(\alpha) e^{|j_4|}\right )^{1/2}
\end{equation}
where $\alpha$ is the ratio of the semi-major axis at the resonance to that of the planet and $f_d(\alpha)$ ranges from 2 to 9 depending on the resonance.  Its values are tabulated in Tab.~8.5 of \citet{solarsystemdynamics} for the resonances of interest.  For the second-order resonances, $|j_4|=2$ and for the first-order resonances, $|j_4|=1$.  Furthermore, for these first-order resonances there are additional terms that increase the width of the resonance that become important for low eccentricity ($e\lesssim 0.1$) \citep[see Fig~8.7 of][]{solarsystemdynamics}.   We can neglect these corrections here because we are only interested in the resonance regions where the orbits of the planetesimals cross that of the planet (above the red line in Fig.~\ref{fig:res_diagram}), i.e. $e>0.2$. The lower-order resonance regions do not overlap for a planet of Neptune mass.  They begin to overlap at the mass of Saturn, so comets that end up in these regions make a sub-dominant contribution.

We now distinguish three cases:
\begin{enumerate}
\item The planet is well within the region evolving adiabatically ($a_p \lesssim 2 \times 10^3$).  This case is the only one appropriate if the bulk of the mass-loss happens during the red-giant branch phase. It also applies to Neptune and the Kuiper Belt objects in our own solar system.
\item The planet is in the adiabatic region but it is close enough to the impulsive region ($2 \times 10^3 \lesssim a_p \lesssim 6\times 10^3$) that many comets can end up on planet-crossing orbits after the mass-loss; either this case or the final one might apply for Planet~9; and 
\item Both the planet and its resonances are in the region that will evolve impulsively ($6 \times 10^3 \lesssim a_p \lesssim 5 \times
  10^4$);
\end{enumerate}
If we assume that the cometary objects are uniformly distributed in $\log a$ and $e^2$ as we do here, the fraction of objects that end up in chaotic regions after the mass-loss increases monotonically from the first case to the third.

In the first case, both the orbits of the comets and the planet will evolve adiabatically, so comets will typically not end up in the orbit crossing regions, but the fraction within the Wisdom-Duncan region will grow by 50\% due the mass-loss from 0.9 to 0.53 solar masses that we consider here or about three times larger for a mass-loss from 2 to 0.6 solar masses, so comets can end up within the chaotic regions due to the expansion of these regions. This growth in the Wisdom-Duncan region will typically bring a fraction of about $6 \times 10^{-5}$ of the planetesimals into chaotic orbits.  Because we will only simulate about $10^4$ planetesimals, we do not expect any of our simulated objects to fall into this chaotic region. Additionally, some comets that are initially in resonance can fall out of resonances because we assume in general that the resonant trapping times are longer than the mass-loss timescale, and the evolution is not strictly adiabatic. In the second case, the planet's semi-major axis is close to the adiabatic-to-impulsive transition region, so comets in the impulsive region will cross the orbit of the planet. Low-eccentricity comets ($e \lesssim 0.1$) will not interact with the planet after the mass-loss except for those initially close the planetary orbit, and only due to the increase in the width of the Wisdom-Duncan region.  In the final case, we expect an higher number of comets to move into the planet-crossing region.

\begin{figure*}
  \includegraphics[height=0.5\textwidth,clip,trim=0.45in 0 0 0]{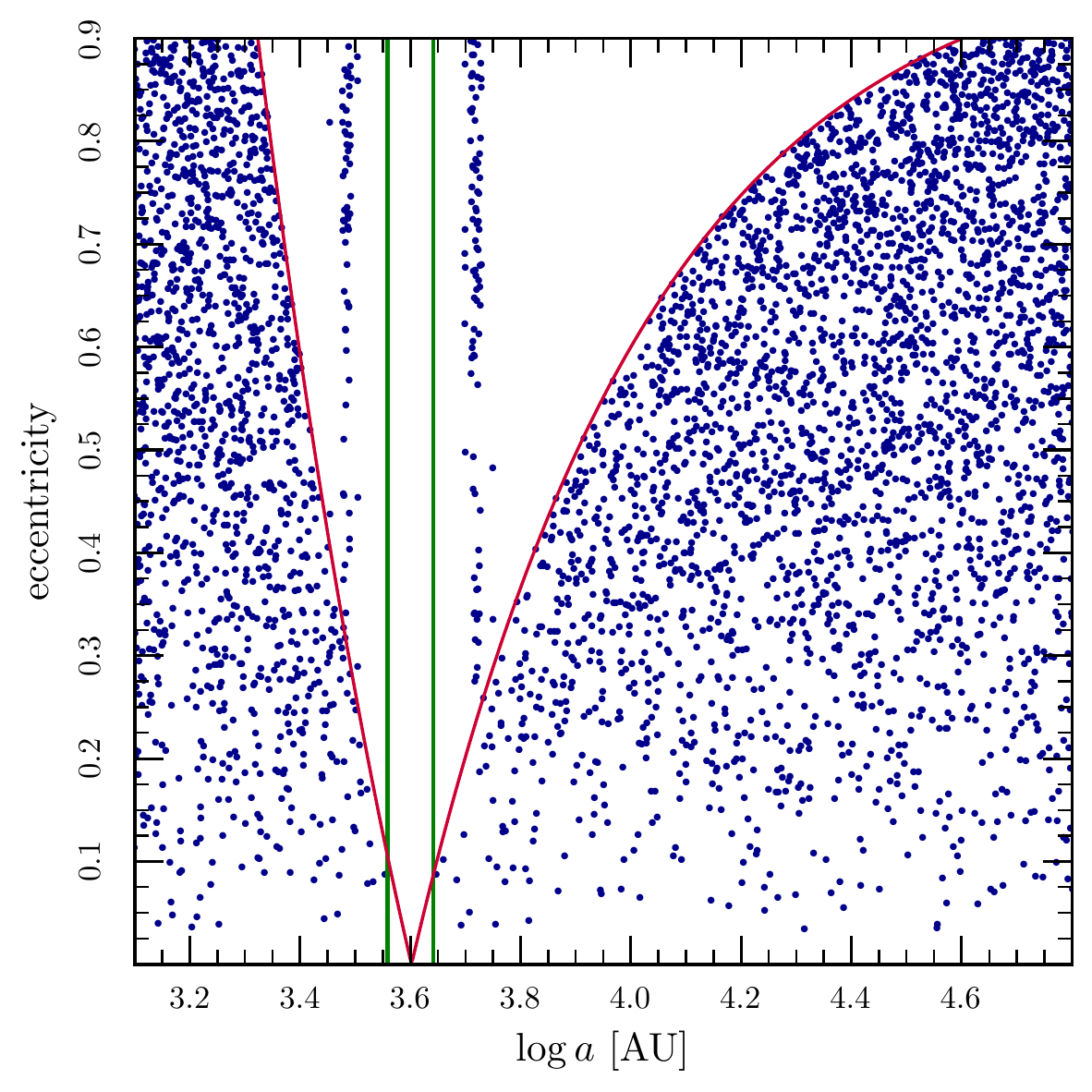}
  \includegraphics[height=0.5\textwidth,clip,trim=0.45in 0 0 0]{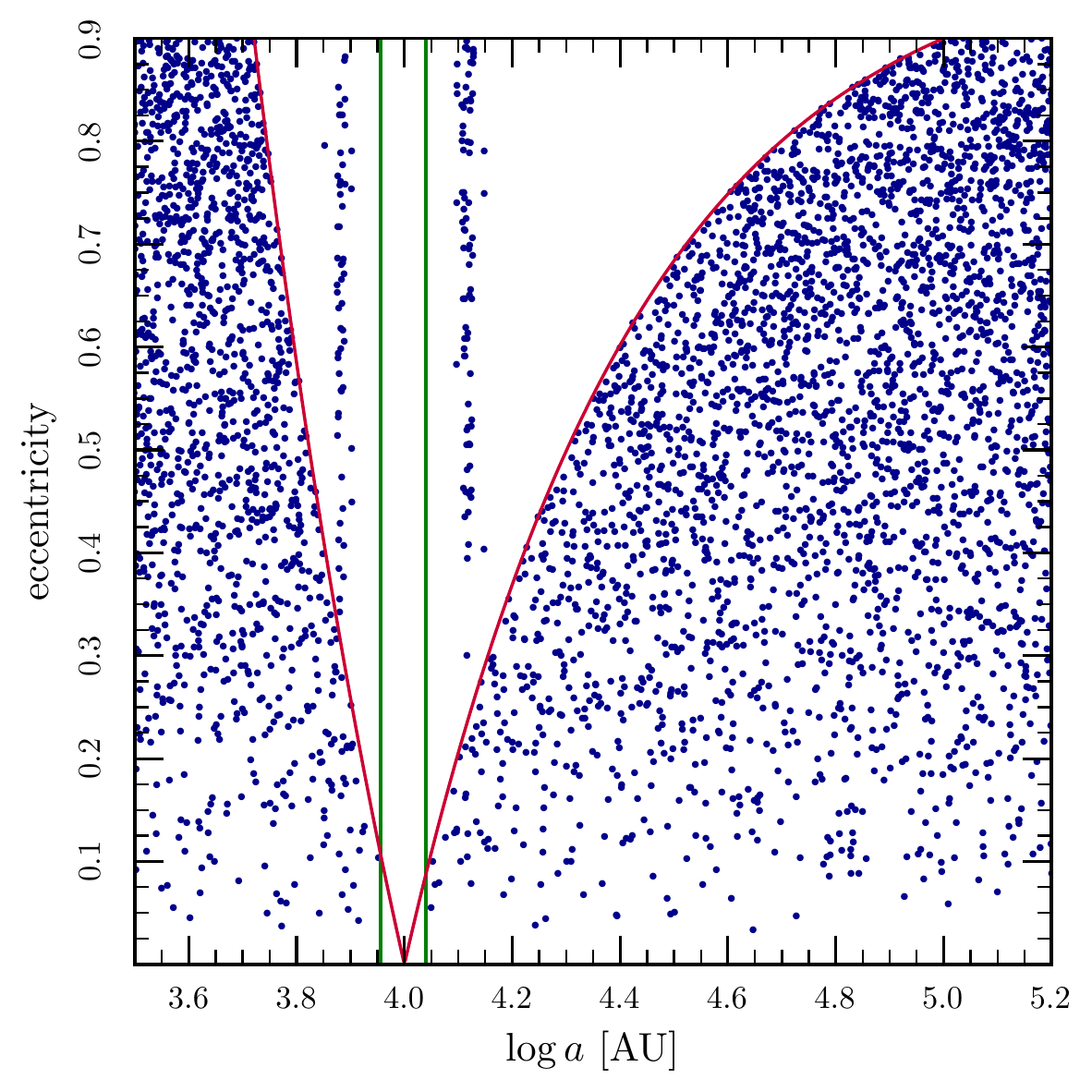}
  \includegraphics[height=0.5\textwidth,clip,trim=0.45in 0 0 0]{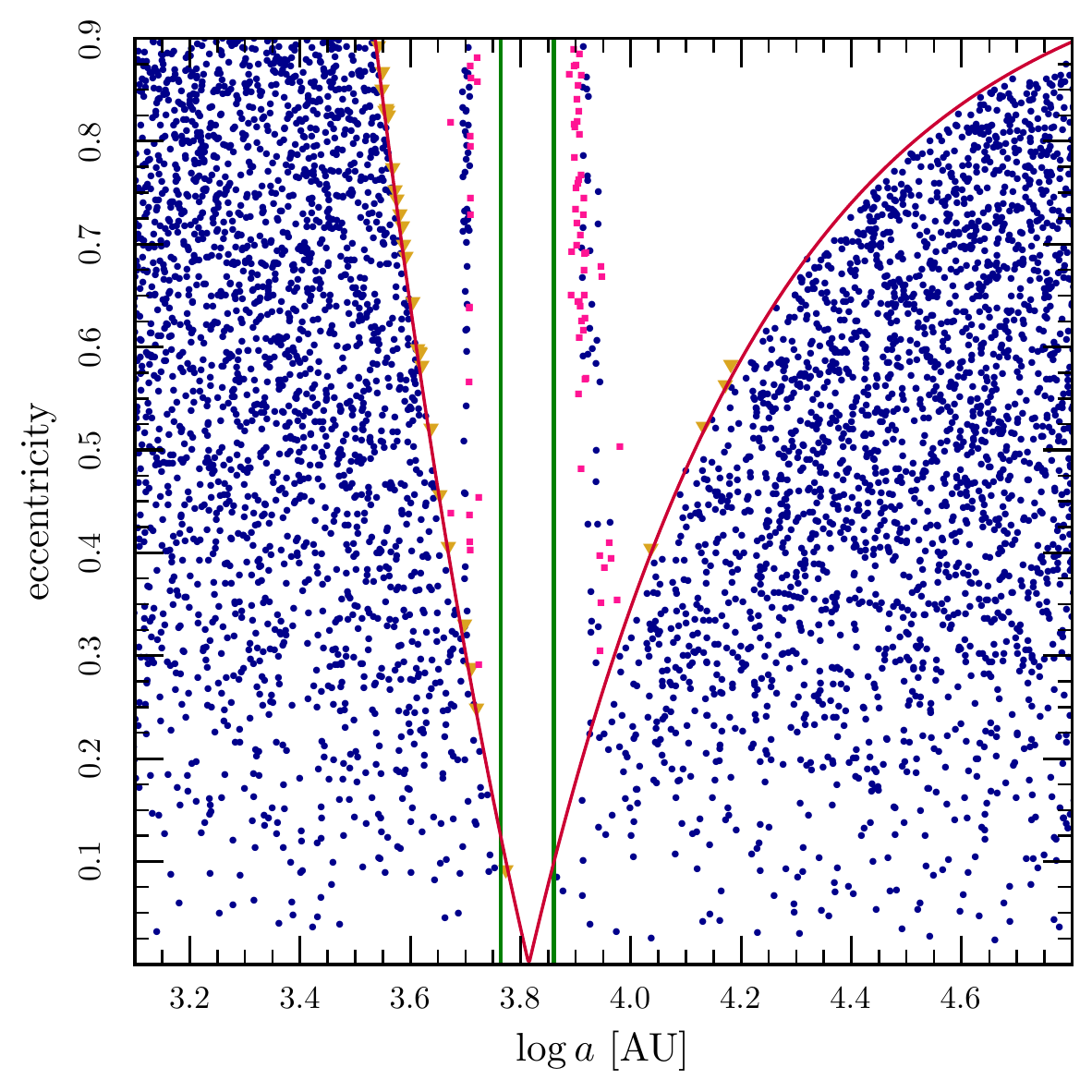}
  \includegraphics[height=0.5\textwidth,clip,trim=0.45in 0 0 0]{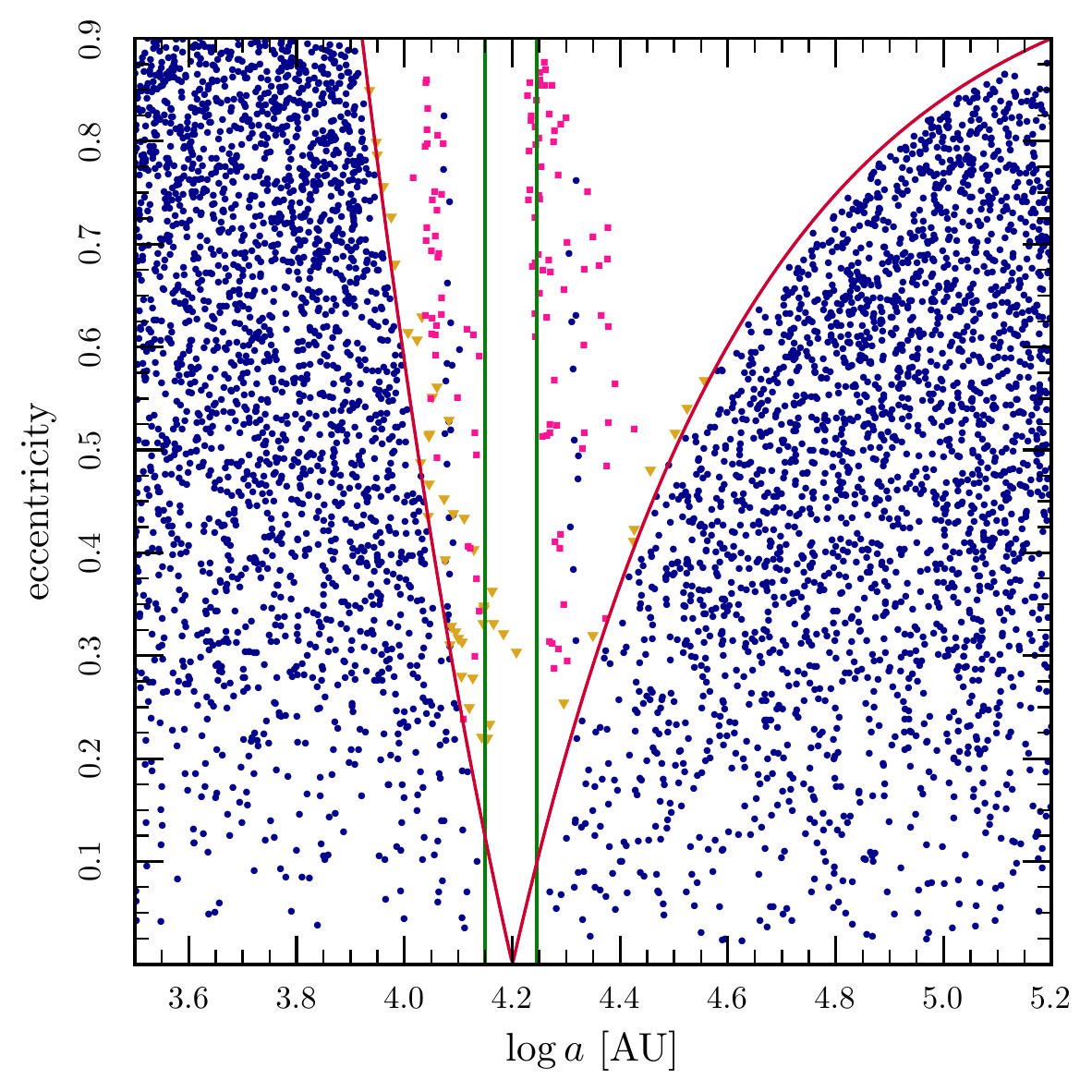}
  \caption{Eccentricity vs logarithm of the semi-major axis for the orbital distribution of the comets before and after the mass-loss for the cyan model.  The red line delineates the orbit crossing region.  The green lines denote the Wisdom-Duncan region.  The two upper panels depict the initial distributions while the lower panels show the final distributions for two different initial positions of the planet. From left to right: left panels, case ii (4,000~AU); right panels, case iii (10,000~AU).  Blue points are comets on stable trajectories.  They either lie within resonances or outside the planet-crossing region.  Pink are comets that initially were in resonance but have fallen out.  Gold are comets that entered the planet-crossing region.  }
    \label{fig:resonance}
\end{figure*}

We simulate the evolution of about 7,500 comets for the cyan mass-loss model (Fig.~\ref{fig:mass-mdot}), which has a long-duration epoch of rapid mass-loss and a high mass on the horizontal branch. Additionally, we run the same simulation for the magenta model of Fig.~\ref{fig:mass-mdot}, which undergoes a more pulsating mass-loss but has a shorter epoch of rapid mass-loss. We pick three different values for the initial semi-major axis for the planet: $10^3$ AU (case i), $4 \times 10^3$ AU (case ii) and $10^4$ AU (case iii). In all cases, we choose the closest object to the selected semi-major axis with low eccentricity among the simulated ones to be the planet. Tab.~\ref{tab:summary} gives the number of comets on initially resonant and non-resonant orbits and the number that end up unstable after the mass-loss for each model. The initial and final distribution of comets are shown in Fig.~\ref{fig:resonance} for case ii and iii of the cyan model. The left panels depict the initial (upper left panel) and final (lower left panel) orbital parameters of the cometary population for a planet initially at $4 \times 10^3$~AU (case ii), while the right panels depict the initial (upper right panel) and final (lower right panel) distribution of comets for a planet initially at $10^4$~AU (case iii). The green lines indicate the Wisdom-Duncan region; the semi-major axis of the planet before and after the mass-loss is indicated by the midpoint between the lines. The red lines delimit the planet's orbit-crossing region. The mean motion resonances are not indicated by lines, but all the stable (blue) comets above the red lines lie in the resonances. Blue dots represent stable comets; pink squares are comets that were initially stable because they were in a resonance and which fell off the resonance during the mass-loss; yellow triangles represent comets that were initially outside of the planet-crossing region and that entered it during the mass-loss phase.

 \begin{table*}
   \caption{Summary of the Orbital Simulations: total number of comets initially and number of unstable comets finally, percentage of initial comets on unstable orbits at the end of the simulation}
   \label{tab:summary}
   \begin{tabular}{ccrrrrr}
     \hline
      Cyan Model&      & \multicolumn{2}{c}{Non-Resonant} & \multicolumn{2}{c}{Resonant} & \\ 
      $\eta_\mathrm{AGB}=0.516$ & $a_p$ [AU]& Total & Unstable  & Total & Unstable & $[\%]$ \\
     \hline
     i   & $1 \times 10^3$ & 7385 &  0 & 141 &  0 & 0 \\
     ii  & $4 \times 10^3$ & 7296 & 30 & 148 & 69 & 1.33 \\
     iii & $1 \times 10^4$ & 7324 & 48 & 137 & 112 & 2.14 \\
     \hline
Magenta Model &       & \multicolumn{2}{c}{Non-Resonant} & \multicolumn{2}{c}{Resonant} & \\ 
$\eta_\mathrm{AGB}=0.21$       & $a_p$ [AU]& Total & Unstable  & Total & Unstable & $[\%]$ \\
     \hline
     i   & $1 \times 10^3$ & 7353 &  2 & 144 &  3 & 0.07 \\
     ii  & $4 \times 10^3$ & 7353 & 20 & 143 & 34 & 0.72 \\
     iii & $1 \times 10^4$ & 7309 & 32 & 145 & 105 & 1.85
   \end{tabular}
 \end{table*}
 
The percentage of comets ending on unstable orbits increases as the initial semi-major axis of the planet gets closer to the impulsive region, as expected. It is also interesting to see the difference in the results between the cyan and the magenta models. In the case of the magenta model, the mass-loss pulses are more intense and, as we can see in Fig.~\ref{fig:mass-mdot}, during pulses the mass-loss rate is comparable with the orbital frequency of objects as close as 3000 AU. For this reason, some of the comets at 1000~AU evolve sufficiently impulsively to end on unstable orbits, while for the cyan model, where the mass-loss rate never reaches high values, we do not see any. On the other hand, the total number of comets that end up on unstable orbits depends on how much mass is lost when  the mass-loss rate is comparable to their orbital frequency. In the magenta model, mass-loss is on average less rapid than in the cyan model, and this results in a smaller number of comets ending up on unstable orbits in cases ii and iii.

\section{The unstable planetesimals}
\label{sec:unstable-planetesimals}

After the orbits of the planetesimals become unstable, they may form a debris disk as in the case of the Helix Nebula \citep[e.g][]{2007ApJ...657L..41S} or fuel late, long-term pollution of the white dwarf atmosphere \citep[e.g][]{2002ApJ...572..556D}.  In the first case, the total mass needed to account for the infrared emission from the debris disks is up to $0.1~\mathrm{M}_\oplus$ \citep{2007ApJ...657L..41S} and it must be delivered within a few million years to about 50~AU from the star.  On the other hand, the mass required for the metal pollution of the atmospheres of older white dwarfs is much more modest, perhaps as little as $10^{-3}~ \mathrm{M}_\oplus$, but it has to be delivered over hundreds of millions of years to a much shorter distance to the star \citep{2002ApJ...572..556D}.

If the planet lies well within the adiabatic region, the fraction of
comets that can be scattered after the mass-loss is quite small and depends on how pulsating the mass loss is.
Just above 7,500 particles were simulated here, so we would not expect any to enter the Wisdom-Duncan region.  Using the size of the Wisdom-Duncan region
(Eq.~\ref{eq:wisdom-duncan}), we find that, for a Neptune mass planet, the
adiabatic increase in the size of this region will result in a few
$10^{-5}$ of the population on chaotic orbits after the mass-loss. This
fraction is proportional to the mass of the planet to the power of
$6/7$.  It also strongly depends on the assumed distribution of
eccentricities.  If the eccentricities are distributed uniformly
instead of $P(e)\propto e$, the fraction increases by a factor of ten
and is proportional to $m_p^{4/7}$.  In both of these cases, this
amount is sufficient to produce a debris disk if the total mass of the
planetesimals exceeds ten Earth masses \citep{2015MNRAS.448..188S}.
The rate at which the comets are scattered by the planet
increases as the mass of the planet 
\citep{2016ApJ...824L..22M,2012A&A...548A.104B} and equal numbers of comets are scattered in successive factors of ten in time \citep{1995Icar..118..322M}.

If the planet lies within the impulsive region or near to it, about
0.6\% of the comets will end up on planet-crossing orbits after the
mass-loss. Therefore, a planet with a much smaller mass than that of Neptune would be sufficient to result in a large fraction of the comets scattered.  We
consider the mass of Mars as a lower limit because, even if its mass is
apparently sufficient to clear its neighbourhood, at least in our solar
system, for such a small mass planet the scattering rate
will also be small \citep{1995Icar..118..322M,2012A&A...548A.104B}, so whether the comets are delivered quickly enough to the inner regions to account for the debris disks and metal lines is not clear. An additional 1.5\% falls out of the resonances and can ultimately be scattered as well. For the most impulsive case iii, this is nearly the entire initially resonant population. To conclude that these comets fall out of the resonances, we have assumed implicitly that the libration timescale is long compared to the mass-loss timescale (see \S~\ref{sec:role-planet}).  Therefore, in this situation, if the total mass of planetesimals exceeded $6~\mathrm{M}_\oplus$, the mass of perturbed planetesimals would exceed $0.1~\mathrm{M}_\oplus$, sufficient to account for the observed debris disks \citep[e.g.][]{2007ApJ...657L..41S} if most of the material can reach about 50~AU where the debris disks lie. 

Both scenarios can explain white dwarfs' metal lines and debris disks depending on the mass of the reservoir of planetesimals in the star system. In our solar system, we have two possible cometary reservoirs: the Kuiper belt, which lies completely in the adiabatic region (for the mass-loss scenarios considered here), and the Oort cloud, which lies at the border between the adiabatic and the impulsive region. According to the current estimates, the mass of the Kuiper belt is less than $0.1~\mathrm{M}_\oplus$, but perhaps an initial mass 100 to 1,000 times larger is required to explain the observed binaries and the abundance of the refractory materials in the giant planets \citep{2006ssu..book..267D}.  What, if anything, eroded the Sun's Kuiper belt is an open question.  However, if the pre-white-dwarf star had an un-eroded belt of planetesimals similar to that the Sun had, this would yield $10-100~\mathrm{M}_\oplus$, in excess of what is needed to explain the formation of debris disks through the dynamical model presented here.  At the upper end of this mass range, even in the case where both the planet and planetesimals lie in the adiabatic region, sufficient planetesimals could be directed inward to form the debris disk.

Estimates of the mass of the Oort cloud are subject to great uncertainties, especially regarding the number of objects in the inner Oort cloud. Current values range from about seven to forty Earth masses \citep{1996ASPC..107..265W,2004ASPC..323..371D,2005ApJ...635.1348F,2008A&A...492..251B}. In our solar system, the Oort cloud is eroded in time due to the interaction with planets, passing stars and giant molecular clouds and due to the Galactic tidal field \citep{1999Icar..137...84W}. The initial mass of the Oort cloud is estimated to have been between 40\% \citep{1987AJ.....94.1330D} and 80\% \citep{1985ASSL..115...87W} larger than what it is now. The Sun, however, lies in a dense region of the galaxy, close to the middle of the Galactic plane; star systems further away from the Galactic plane may experience a less dramatic erosion of their cometary population due to fewer encounter with passing stars and molecular clouds. Moreover, the progenitors of the polluted white dwarfs that we observe lived a shorter life compared to our Sun, and so they are likely to have retained more comets than our Sun has up to its current age.

\section{Conclusions}
\label{sec:conclusions}
We present a model with realistic mass-loss on the AGB and a heuristic
approach to the orbital evolution of the exo-Oort cloud and exo-Kuiper belt to calculate
the evolution of a population of exo-comets as a white dwarf is
formed.  If the cometary population has a planetary mass within it, as
there are hints in our own solar system \citep{2016AJ....151...22B},
comets initially on stable orbits can end up on chaotic orbits either
due to the expansion of the chaotic regions \citep[this expansion was first
pointed out by][]{2002ApJ...572..556D} or to impulsive evolution, and
can be directed toward the inner stellar system.

An important subtlety here is that, if the mass-loss occurs on the red-giant branch, the comets can only end up in the chaotic regions due to the expansion of these regions. More importantly, in this case the comets fall into the chaotic regions while the star is still a red giant. Thus, it is most likely that they will end up in the inner regions during the horizontal branch phase, so the resulting debris disk will be destroyed as the star ascends the AGB.

We have not discussed here the formation of the debris disk and how precisely metals are supplied to white-dwarf atmospheres.
\citet{2015MNRAS.448..188S} explored these issues in detail, and their
model would apply here as well.  The primary difference is how the
comets end up on orbits that plunge close enough to the star that they
sublimate.  \citet{2015MNRAS.448..188S} rely on the white-dwarf
getting a kick during the mass-loss, while we assume that a planet of
at least the mass of Mars lies at a similar semi-major axis to the
comets.  The key missing ingredient in the model here that will
require more detailed simulations is an estimate of the fraction of
comets in the chaotic regions or on planet-crossing orbits that are
scattered toward in the innermost regions of the system.  The
requirements are not terribly stringent because a large fraction
of the comets can end up on these perturbed orbits.  In our solar
system, the closer-in planets scatter many of the objects that come
close enough to the star to sublimate. In the evolutionary path calculated in this paper, however, due to the fact that during the RGB the star does not lose much mass, a star like our Sun would expand to large radii during the AGB stage, vaporizing the planets within about 5~AU and thus clearing the inner stellar system of additional scatterers \citep[see also][]{2002ApJ...572..556D}.

In the paper we considered two possible reservoirs of planetesimals: the equivalents of our Kuiper belt and our Oort cloud. We showed that both types of reservoir can supply the amount of comets needed to pollute white-dwarf atmospheres and create debris disks if a planet is present at a similar semi-major axis, with the Oort cloud being a better candidate because it lies at the border between the adiabatic and the impulsive region and because it has a bigger cometary mass. We did not consider the possibility of the polluting planetesimals coming from an inner region like our asteroid belt. In the solar system, the asteroid belt is shaped by the interaction of mean-motion resonances and secular resonances \citep[e.g][]{solarsystemdynamics}.  An analysis of the asteroid belt's dynamics is well beyond the scope of this paper and would probably not be applicable to other stellar systems. However, the presence of a planet shepherding these inner planetesimals is likely to be a robust feature. In that case, as the star loses mass, the Wisdom-Duncan region of the planet grows and perhaps other resonances overlap if the mass of the planet is at least that of Saturn.  This would cause a large fraction, perhaps as large as $10^{-2}$ for a Jupiter-mass planet, of the planetesimals in the belt to end on chaotic orbits after the mass-loss. If the mass of planetesimals in the inner region exceeded 0.1~M$_\oplus$, this scenario would be sufficient to account for the formation of a debris disk. This is not the case for our solar system, where the total mass of the asteroid belt is only $5\times 10^{-4}\mathrm{M}_\oplus$.

In conclusion, the final rapid mass-loss of a star can restructure the system of planets and planetesimals surrounding it.  If the regions of planetesimals in the stellar system are essentially bounded in phase space by the influence of planets (as in our solar system), the expansion of the chaotic regions into formerly stable regions can fuel a late and somewhat heavy bombardment of the inner stellar system.  On the other hand, if a planet lies further out, approaching or within the region when orbits respond impulsively (e.g. Planet~9), the scrambling of the cometary orbits and the planetary orbit will result in the scattering of an even larger fraction of the planetesimals.  In all these cases, the mass of the planetesimals scattered can be sufficient to account for the debris disks inferred around white dwarfs and for the metal lines in their spectra.

{\noindent \bf Acknowledgements}

We would like to thank the anonymous referee for detailed comments and constructive suggestions that we feel improved the paper dramatically.
We used the NASA ADS service and arXiv.org. This work was supported by
the Natural Sciences and Engineering Research Council of Canada, the
Canadian Foundation for Innovation and the British Columbia Knowledge
Development Fund.

\bibliography{wd_pollution} \bibliographystyle{mnras}

\label{lastpage}
\end{document}